\documentclass[%
 reprint,
unsortedaddress,
nofootinbib,
 amsmath,amssymb,
 aps,
 prx,
floatfix,
longbibliography]{revtex4-2}
\makeatletter
\renewcommand\frontmatter@abstractwidth{\dimexpr\textwidth\relax}

\makeatother
\usepackage{comment}
\usepackage{graphicx}
\usepackage{dcolumn}
\usepackage{bm}
\usepackage[dvipsnames]{xcolor}
\usepackage{mathrsfs}

\usepackage{graphicx}
\usepackage{dcolumn}
\usepackage{bm}
\usepackage[utf8]{inputenc}
\usepackage[T1]{fontenc}
\usepackage{mathptmx}
\usepackage{etoolbox}
\usepackage{amsmath}
\usepackage{caption}
\usepackage{svg}
\usepackage{float}
\usepackage[normalem]{ulem}
\usepackage[dvipsnames]{xcolor}
\usepackage{amssymb}

\begin{document}

\preprint{APS/123-QED}

\title{Spatiotemporal Disk Packing for Directed Growth of Complex Geometries}

\author{Yigit Hergul, Yun Seong Kim, Rohan Shah, Sameh Tawfick, and Varda F. Hagh}
\email[To whom correspondence should be addressed: ]{hagh@illinois.edu}
\affiliation{%
Mechanical Science and Engineering, and Beckman Institute for Advanced Science and Technology,
University of Illinois at Urbana-Champaign, Urbana, Illinois 61801, USA}%

\begin{abstract}
Growing complex shapes requires control over both where growth begins and how it evolves in time. Here, we introduce a geometric framework for growing prescribed 2D shapes using a disk packing algorithm. In this approach, a target geometry is filled by disks whose centers define where growth is initiated and whose radii define how long each region is allowed to grow. The allowed disk sizes are constrained by the physics of the process, including the growth velocity, the time required to initiate each growth event, and the number of initiations that can occur in parallel. To generate physically realizable packings, we introduce the Largest Gap Algorithm (LGA), which sequentially fills the largest remaining gaps in a target shape with the largest disk that satisfies both geometric and kinetic constraints. We show that this method produces high coverage packings for a variety of geometries and that the resulting packings can be directly converted into spatiotemporal packing instructions. We then demonstrate that these instructions can be realized experimentally using multi-point initiation of frontal polymerization in viscosified dicyclopentadiene (DCPD) resin using CO$_2$ laser. Our results show that complex shapes can be grown by programming a small number of local initiation events, providing a simple connection between geometry and dynamics of growth.  
\end{abstract}

\maketitle

\section{Introduction}

Many physical and biological systems generate complex shapes and structures through localized nucleation followed by outward growth. Examples include the expansion of supercritical crystal nuclei in supersaturated solutions~\cite{langer1980instabilities,lee2016multiple,escobar2021direct}, solidification of reaction-diffusion fronts in reactive media~\cite{chernavskii1991model,yan2006frontal,itatani2018role,kim2025morphogenic}, spreading of bacterial colonies across nutrient substrates~\cite{ben1994generic,dayel2009silico,narla2021traveling}, and the sequential initiation and outward displacement of plant primordia during phyllotaxis~\cite{douady1992phyllotaxis,fleming2005formation,coen2024developmental}. Despite their differences, these systems can be viewed through a common geometric lens in which localized domains originate at discrete sites, grow or are displaced outward, and eventually interact with neighboring domains before arresting under spatial constraints. This shared geometry motivates a broader view of shape formation as a process governed by the spatial arrangement and temporal sequence of multiple growth events.

Several theoretical frameworks have described such growth processes as packings of expanding disks or spheres. In touch-and-stop~\cite{andrienko1994pattern,tsakiris2011percolation,chiarelli2015stochastic,aboufoul2019virtual} and lilypond~\cite{haggstrom1996nearest,cotar2004note,hirsch2015stationary} models, as well as in related packing-limited growth processes~\cite{dodds2002packing,dodds2003packing,auclair2023mean}, domains grow from seeded locations and stop when they encounter neighboring domains.  These models provide a natural forward description in which rules governing local nucleation and growth determine the global structure of the final packing. We use the same geometric representation in a complementary direction. Rather than predicting the structure that emerges from specified nucleation sites and growth rules, we ask how a prescribed target geometry can be generated by intentional positioning and temporal sequencing of growing domains.

This inverse question has a simple geometric interpretation. A $2$D target geometry can be approximated by placing $n$ disks so that they cover the desired area, then initiating growth at the center of each disk and allowing them to expand to their final radius. In this representation, the number of disks, their positions, and their radii are the degrees of freedom that must be determined~\cite{hagh2022transient}.
Without additional constraints, however, this coverage problem is under-determined. 
For instance, one could use arbitrarily many small disks, take $n \to \infty$, and approximate the target geometry with arbitrary accuracy. One could also begin with the largest inscribed disk and recursively fill the remaining gaps with smaller disks, producing an Apollonian-like hierarchy of sizes~\cite{kasner1943apollonian,varrato2011apollonian}. Alternatively, if overlaps among disks are allowed, a finite number of disks can be used to cover the target with high accuracy. 

These examples show that the target geometry alone does not define a unique packing strategy. Without physical constraints, there are infinitely many ways to cover a target shape with disks whose number, positions, and sizes are freely chosen. In many growth processes, however, these quantities are constrained by the kinetics of nucleation and propagation. The relevant packing problem is therefore not to find a random disk packing that covers the target geometry, but to find one that can be physically realized by a given growth process.

To construct such physically accessible packings, we introduce a sequential packing algorithm that recursively fills the largest voids using only disks that can be realized by the growth process. The construction begins with the largest disk that fits within the target geometry and simultaneously satisfies the kinetic constraints. The remaining gaps are then identified and filled iteratively with smaller accessible disks. In this way, the algorithm constrains both the positions and sizes of the disks, by assigning larger disks to wider parts of the geometry and smaller disks to progressively finer features, while also respecting both the geometric constraints imposed by the target shape and the physical constraints imposed by the kinetics of growth.

We demonstrate that this algorithm can program the growth of diverse target geometries with high geometric fidelity and show that the resulting growth sequences can be realized experimentally using multi-point frontal polymerization of dicyclopentadiene (DCPD)~\cite{kim2025morphogenic}. Frontal polymerization is an exothermic, self-sustaining curing reaction in which a monomer is locally heated above the temperature required to initiate polymerization, after which a reaction-diffusion front propagates \textit{isotropically} through the medium~\cite{suslick2023frontal,malik2020review,goli2020frontal}. When front propagation is isotropic, each locally initiated front expands radially from its nucleation site, forming circular growth domains in $2$D and spherical growth domains in $3$D.

By initiating fronts at multiple locations and times using an appropriate path-planning algorithm, this approach can realize target geometries that are well approximated by disk packings in $2$D or sphere packings in $3$D. Because the reaction front in frontal polymerization propagates rapidly, this method offers a rapid growth-based manufacturing strategy for producing near-net complex shapes. Here we focus on $2$D target geometries, but the same framework can be extended to $3$D by replacing disks with spheres and area coverage with volume filling.

\section*{Methodology}
We model each locally initiated growing front as a disk. The center of the disk corresponds to the nucleation site, and its radius is determined by the amount of time available for growth. The boundary of the target geometry is treated as a hard constraint, so each disk must remain fully inside the target shape. 
The process constraints are set by the front velocity $v_{\rm f}$, the dwell time $t_{\rm d}$, and the number of available initiators $N_{\rm i}$. The front velocity determines how far a front can propagate during a given amount of time. The dwell time is the time required to initiate nucleation at a site by providing heat as energy input. During this period, the corresponding initiator is occupied and cannot initiate growth elsewhere. In practice, the initiator requires a finite amount of time to travel between consecutive nucleation sites, but this travel time is assumed to be negligible here. Finally, the number of initiators limits how many fronts can be initiated simultaneously.
\begin{figure}[h]
\centering
\includegraphics[width=1\linewidth]{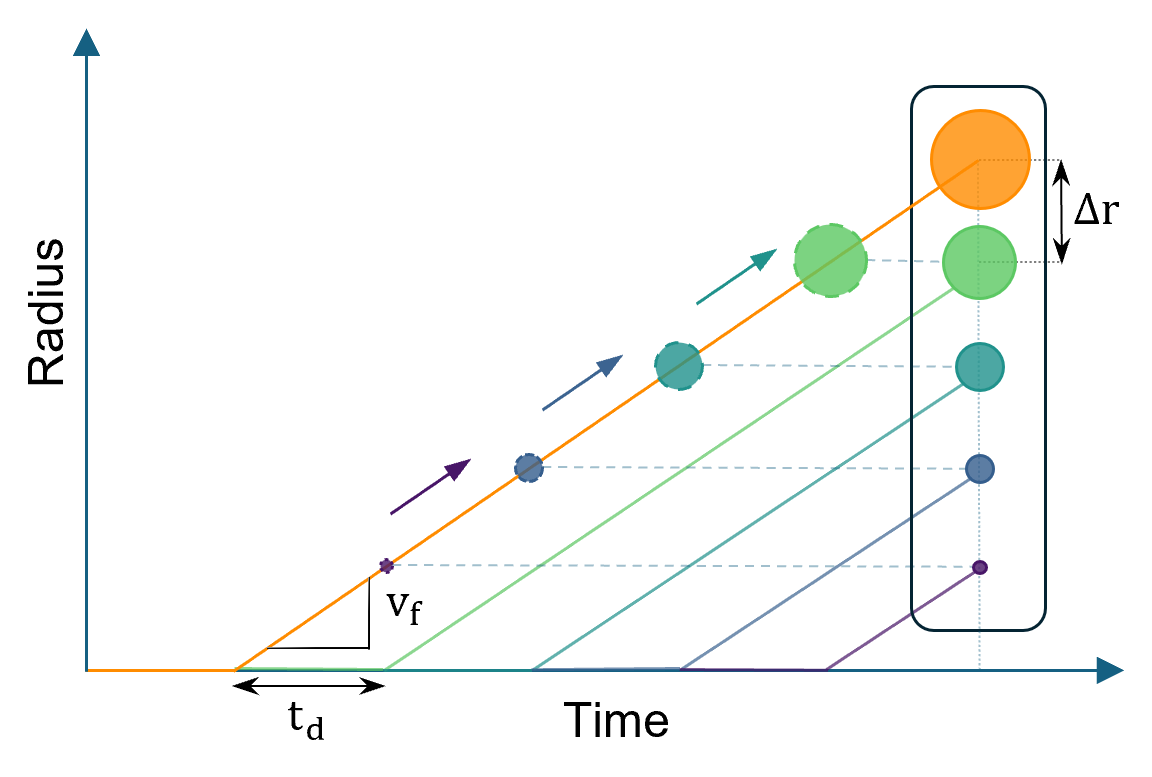}
\caption{Discrete disk sizes generated by initiating growth at different times. Each nucleation site grows isotropically at a constant radial velocity after initiation. Because later nucleation events have less time to grow, they produce disks with smaller final radii. The time interval between consecutive initiations is the dwell time which leads to a discrete set of disk sizes available for packing the target geometry.}
\label{fig:discrete_sizes}
\end{figure}

These constraints restrict the set of disk radii that can be realized. For a single initiator, consecutive fronts cannot be initiated at arbitrarily close times because each initiation requires a dwell time. The minimum spacing between the radii of two consecutive disks is $\Delta r = v_{\rm f} t_{\rm d}$. This spacing arises because while the initiator heats the next site for a time $t_{\rm d}$, the previously initiated front continues to grow by $v_{\rm f} t_{\rm d}$. Repeating this process generates a discrete set of accessible radii. This construction is illustrated in Fig.~\ref{fig:discrete_sizes}, where each nucleation site grows with slope $v_{\rm f}$ after initiation and reaches its final radius at the end of the growth period. 
\begin{figure}[h]
\centering
\includegraphics[width=0.95\linewidth]{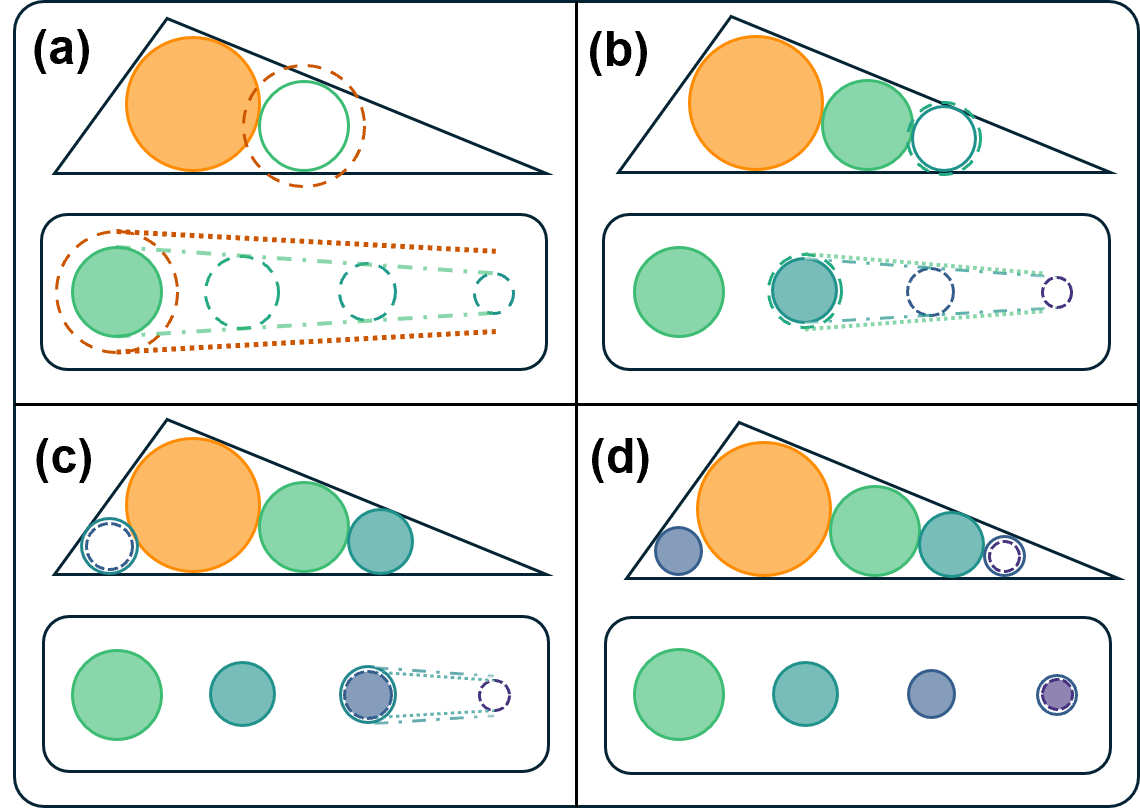}
\caption{ Visual description of the Largest Gap Algorithm (LGA) without overlaps. 
(a) After placing the largest inscribed disk, the largest remaining gap is identified. Two tentative disks are considered at the center of this gap: one set by the geometry of the gap (solid line) and the other by the next kinetically accessible disk size (dashed line). The smaller of these two determines the admissible disk, and the list of available disks is updated accordingly. The dashed, unfilled disk in the box denotes the kinetic constraint for subsequent placements. The dash-dotted and round-dotted lines indicate the future kinetic limits of consecutive disks for the disks previously constrained by geometry and kinetics, respectively.
(b) The chosen disk, shown with infill, is placed in the largest gap. (c, d) When the kinetic constraint is dominant, the successive candidate disks retain the discrete accessible sizes imposed by the kinetics. The process ends when either the smallest remaining gap or the next accessible disk size falls below the minimum allowed radius.
}
\label{fig:LGA}
\end{figure}
The value of the largest radius is limited by the largest inscribed disk that fits within the target geometry. The smallest radius is set by the smallest growth interval that can be resolved. Here, we take $t_{\rm d}$ as the minimum time interval for growth and treat shorter intervals as inaccessible.

Given this discrete set of accessible radii, we next determine where disks of these sizes should be placed within the target geometry. To do so, we use a sequential packing procedure, referred to here as the Largest Gap Algorithm (LGA), as illustrated in Fig.~\ref{fig:LGA}. The algorithm begins by identifying the largest unfilled region within the target boundary and placing the largest accessible disk that can fit in that region. After each placement, the occupied region is updated, and the largest remaining gap is identified. The algorithm then fills this gap with the closest available disk size from the discrete radius set, while satisfying the geometric constraints. This process is repeated until the target area is covered or until no additional accessible disk can be placed. 

A central challenge is that the size of the remaining gap is often smaller than the next largest accessible disk radius. As a result, a strictly non-overlapping packing would leave large uncovered regions and introduce longer idle periods for the initiators. To address this mismatch, the LGA allows controlled overlap between neighboring disks. This overlap provides the flexibility needed to use the available discrete disk sizes while maintaining high coverage of the target area.

Once all possible disks have been placed within the target boundary, we reduce unnecessary overlaps by introducing a repulsive interaction potential between disks. For disks $i$ and $j$, the interaction potential energy is given by $U_{ij}= \epsilon_{ij} \big( \lvert \bm{X}_i - \bm{X}_j \rvert - [R_i + R_j] \big)^2 \Theta\big( \lvert \bm{X}_i - \bm{X}_j \rvert - [R_i + R_j] \big)$, where $\epsilon_{ij}$ sets the energy scale, $\bm{X}_i$ is the position of disk $i$, $R_i$ is its radius, and $\Theta$ is the Heaviside step function introduced to ensure that the interaction is active only when two disks overlap. The target perimeter is treated as a hard boundary, preventing disks from moving outside the prescribed shape. We use steepest descent to minimize the total energy, defined as the sum of all pairwise interactions. The minimization redistributes the disks within the target boundary and reduces their overlap while preserving the initiation sequence generated by the algorithm.

The resulting disk packing defines a realizable growth plan for the target geometry. The center of each disk specifies where a nucleation should be initiated, while its radius determines when that initiation should occur within the growth sequence. Assuming a constant front velocity, $v_{\rm f}$, larger disks must be initiated earlier because they require more time to reach their final radii. The packing can therefore be translated directly into a spatiotemporal plan specifying where and when each growth event should be initiated.

\section*{Results and Discussion}

\textit{Geometric Fidelity}. We use the Largest Gap Algorithm (LGA) to pack disks within different target shapes and quantify how well the resulting packing approximates each geometry using the area coverage ratio, $A_{\rm r}$. This ratio is defined as the area covered by the disks divided by the total area of the target geometry.  
\begin{figure}[h]
\centering
\includegraphics[width=1\linewidth]{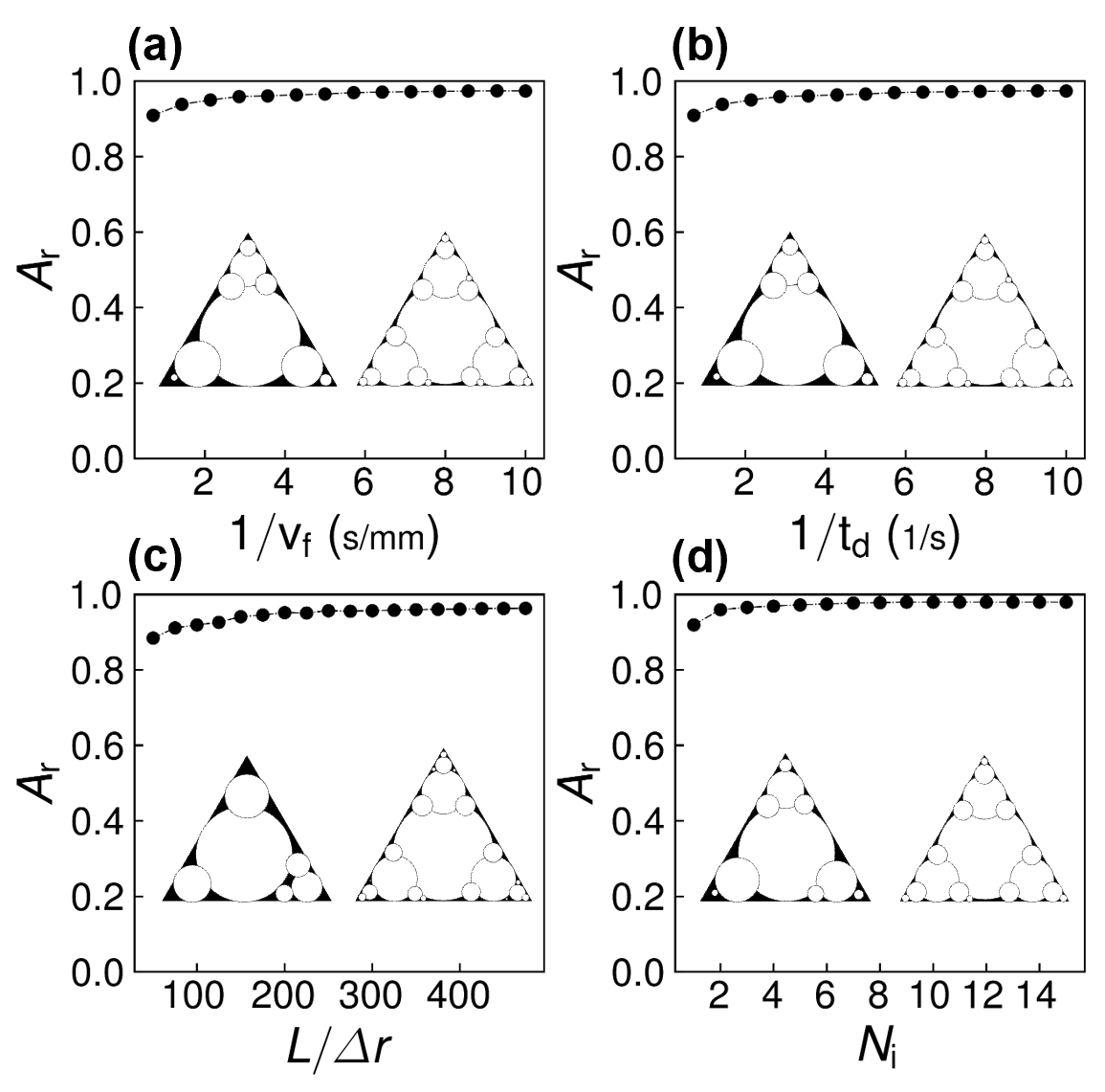}
\caption{The covered area is evaluated as a function of (a) front velocity $v_{\rm f}$, (b) dwell time $t_{\rm d}$, (c) the normalized target size $L/\Delta r$, and (d) the number of independent initiators $N_{\rm i}$. Circle packings in equilateral triangles are shown as insets for the first and last data points of each plot, on the left and right, respectively.
The spacing between consecutive accessible disk radii is $\Delta r = v_{\rm f} t_{\rm d}$, so decreasing $v_{\rm f}$ or $t_{\rm d}$, or increasing $L/\Delta r$, increases the number of accessible disk sizes and generally improves coverage.  Increasing $N_{\rm i}$ also improves coverage by allowing more initiation events to occur in parallel. The fixed parameters for panels are $v_{\rm f}=1~\mathrm{mm/s}$, $t_{\rm d}=1~\mathrm{s}$, $N_{\rm i}=1$, and $L=100~\mathrm{mm}$. }
\label{fig:MyFig4_LGA_Performance}
\end{figure}
Values of $A_{\rm r}$ approaching unity indicate that the disk packing covers the surface area of the target shape closely. We first examine an equilateral triangle with side length $L$ and then extend the analysis to other geometries. The triangle provides a suitable case study because its acute corners are difficult to fill with circular fronts, making it useful for evaluating the limitations of our packing strategy.
\begin{figure}[h]
\centering
\includegraphics[width=1\linewidth]{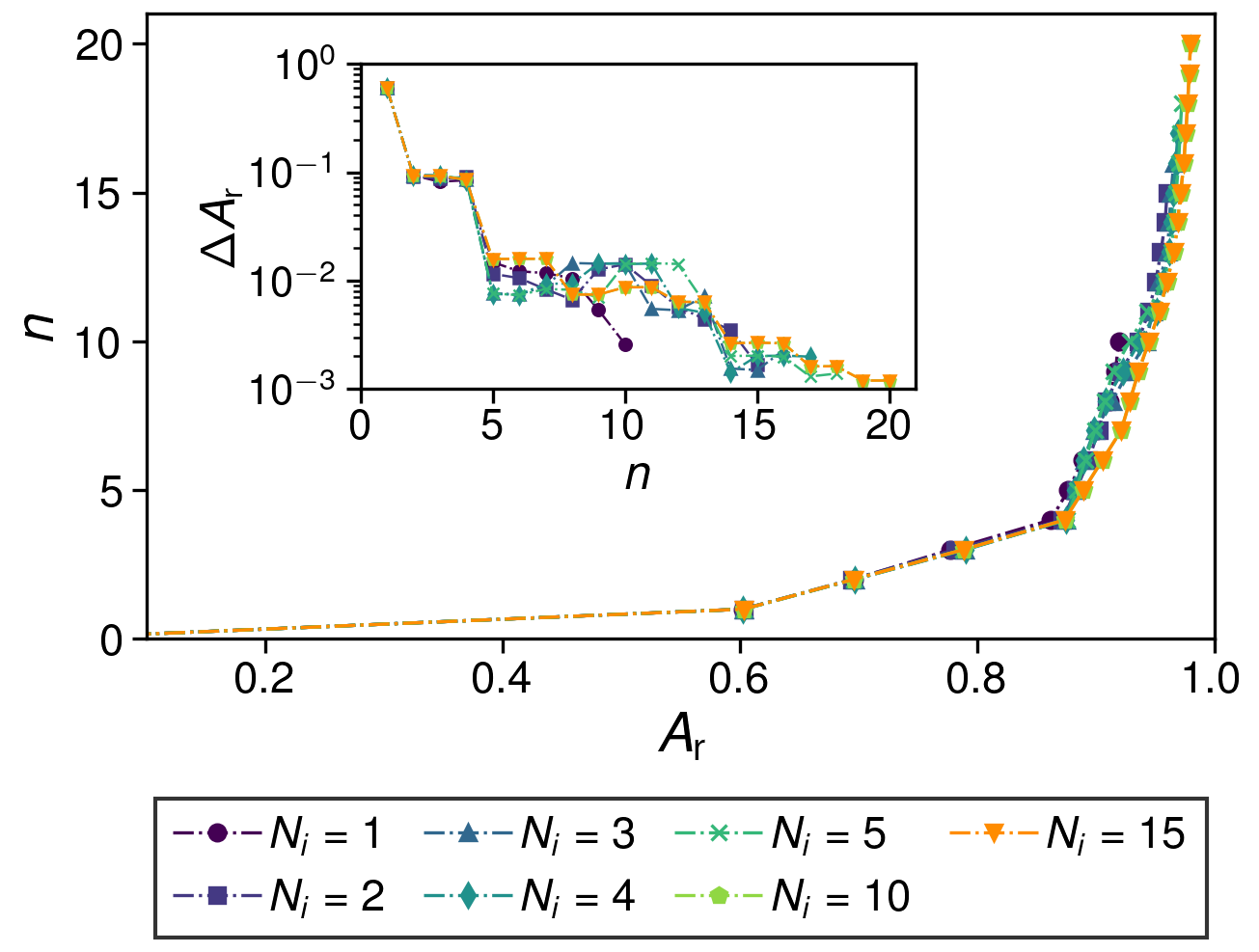}
\caption{Number of initiated disks as a function of area coverage ratio for packings generated by the LGA with different numbers of parallel initiators, $N_i$. Because each disk corresponds to one nucleation event, $n$ is proportional to the total initiation energy for fixed initiation conditions. The inset shows incremental gain in coverage, $\Delta A_{\rm r}(n)$, as disks are added sequentially by the LGA. The fixed parameters are $v_{\rm f}=1~\mathrm{mm/s}$, $t_{\rm d}=1~\mathrm{s}$, and $L=100~\mathrm{mm}$.
}
\label{fig:energy-cost}
\end{figure}

The area coverage ratio depends on the main process parameters, including the front velocity $v_{\rm f}$, the dwell time $t_{\rm d}$, and the number of independent initiators $N_{\rm i}$. Increasing either $v_{\rm f}$ or $t_{\rm d}$ increases the spacing between accessible disk radii, $\Delta r = v_{\rm f} t_{\rm d}$. This reduces the number of available disk sizes for a given target geometry and lowers the likelihood of achieving high coverage. Increasing the number of initiators $N_{\rm i}$, on the other hand, improves coverage by providing greater flexibility in the initiation sequence. With more initiators, multiple fronts can be initiated in parallel, allowing more disks with similar radii to be used. This relaxes the constraints imposed by the discrete radius sequence and enables the algorithm to fill the target geometry more effectively.
Another important factor is the geometric ratio between the characteristic length of the target shape, $L$, and the process-defined length scale $\Delta r = v_{\rm f} t_{\rm d}$.  This ratio controls the number of accessible disk sizes. In our construction, the smallest accessible radius is taken to be $\Delta r$, while the largest accessible radius is limited by the largest disk that can fit within the target geometry and therefore is bounded by $L$. Because consecutive accessible radii are separated by $\Delta r$, increasing $L/\Delta r$ increases the number of sizes available for packing. A larger $L/\Delta r$ allows the target geometry to be resolved more finely and covered more effectively, whereas a smaller $L/\Delta r$ restricts the available radii and reduces the coverage efficiency.  Fig.~\ref{fig:MyFig4_LGA_Performance} shows how $A_{\rm r}$ varies with $1/v_{\rm f}$, $1/t_{\rm d}$, $N_{\rm i}$, and $L/\Delta r$.

\textit{Energy Cost}. In a physical growth process, each nucleation event has an energetic cost. For instance, in frontal polymerization, the material must be heated locally above an initiation threshold before the reaction front can propagate autonomously. If the energy required to initiate each front is approximately fixed, then the total initiation energy is proportional to the number of disks, $n$. So $n$ provides a simple measure of the energetic cost of a packing. In Fig.~\ref{fig:energy-cost}, we plot $n$ as a function of the area coverage ratio, $A_{\rm r}$ for different values of $N_{\rm i}$, showing how many initiation events are needed to achieve a given area coverage in each case. The inset depicts the increase in the area coverage ratio, $\Delta A_{\rm r}(n)=A_{\rm r}(n)-A_{\rm r}(n-1)$, as disks are added one at a time by the algorithm. In the absence of any disks, $A_{\rm r}=0$, whereas the first disk produces a large increase in coverage $\Delta A_{\rm r}\approx 0.60$. Subsequent disks progressively fill the remaining gaps, but the marginal increase in the area coverage decreases and eventually becomes incremental as the target becomes more highly covered. While the gain in coverage, $\Delta A_{\rm r}$, approaches zero, each added disk still requires a separate initiation event and therefore an additional energy input. These plots make the tradeoff between geometric fidelity and energetic cost explicit, since beyond some coverage threshold, the gain in $A_{\rm r}$ may be too small to justify another initiation event. The energetically favorable packing may therefore stop before the smallest remaining gaps are filled.

\begin{figure}[h]
\centering
\includegraphics[width=1\linewidth]{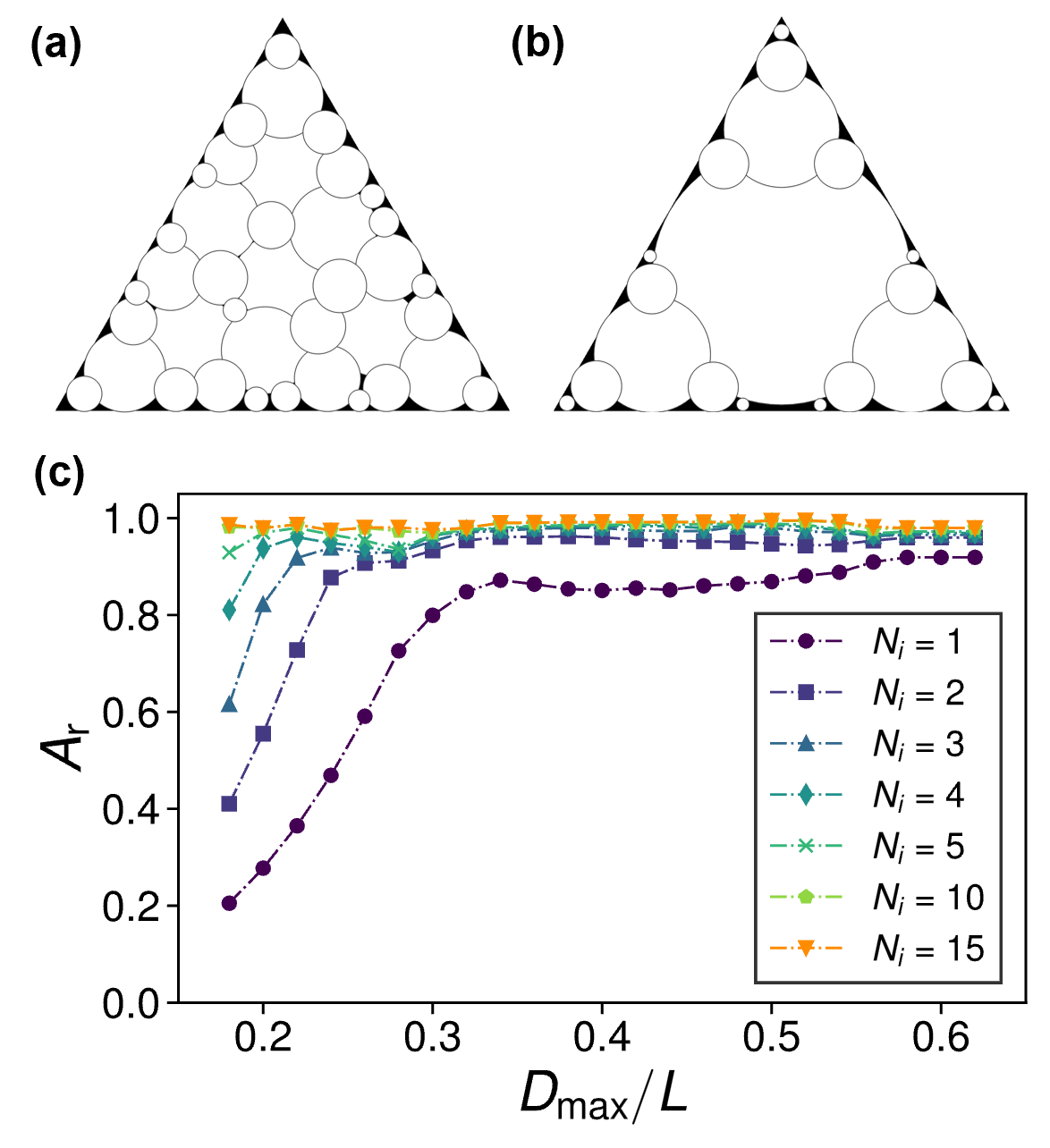}
\caption{(a,b) Disk packings generated with different values of the normalized maximum diameter $D_{\rm max}/L$ for $N_{\rm i}=15$, both achieving an area coverage ratio of $A_{\rm r}=0.98$. 
The packing in (a) uses the smallest value of $D_{\rm max}/L = 0.20$ that reaches this coverage, whereas the packing in (b) uses the largest possible value of $0.58$, corresponding to the largest inscribed disk. 
(c) Area coverage ratio as a function of $D_{\rm max}/L$ for different numbers of initiators $N_{\rm i}$. The ratio $D_{\rm max}/L$ provides a size-independent parameter for comparing packings across target sizes. The fixed parameters are $v_{\rm f}=1~\mathrm{mm/s}$, $t_{\rm d}=1~\mathrm{s}$, and $L=100~\mathrm{mm}$. }
\label{fig:MyFig6_Ni}
\end{figure}

\textit{Growth Time}. We note that the total time required to grow a target shape is set by the radius of the largest disk in the packing, $R_{\rm max}$. This disk is initiated first, and the growth continues until it reaches its final size. However, choosing the largest inscribed disk as $R_{\rm max}$ does not always maximize the area coverage ratio, because it fixes the accessible radius sequence in increments of $\Delta r$, and that sequence may not match the sizes of the remaining gaps in the target geometry. In some cases, selecting a smaller value of $R_{\rm max}$ can both improve the accuracy of the packing and reduce the total time required for growth. Even when the area coverage ratio remains unchanged, as in Figs.~\ref{fig:MyFig6_Ni}a and~\ref{fig:MyFig6_Ni}b, choosing a smaller maximum disk can substantially accelerate the process. For the packing in Fig.~\ref{fig:MyFig6_Ni}a, the same area coverage, $A_{\rm r}=0.98$, is achieved in nearly one third of the time required for the packing based on the largest inscribed disk in Fig.~\ref{fig:MyFig6_Ni}b.

Fig.~\ref{fig:MyFig6_Ni}c shows how the area coverage ratio varies with the dimensionless quantity $D_{\rm max}/L$ for different numbers of initiators, $N_{\rm i}$. Here, $D_{\rm max}=2R_{\rm max}$ is the diameter of the largest disk. We use the diameter so that $D_{\rm max}/L$ lies between $0$ and $1$. 
For a given front velocity and dwell time, plots such as those in Fig.~\ref{fig:MyFig6_Ni}c can be used to identify the smallest value of $D_{\rm max}/L$ that achieves a desired area coverage ratio. For very small values of $D_{\rm max}/L$, the algorithm performs poorly because the largest available disk is too small. In this regime, the fixed radius spacing $\Delta r$ limits the number of accessible disk sizes, causing the algorithm to run out of usable disks before covering a sufficient fraction of the target area.

\begin{figure}[h]
\centering
\includegraphics[width=1\linewidth]{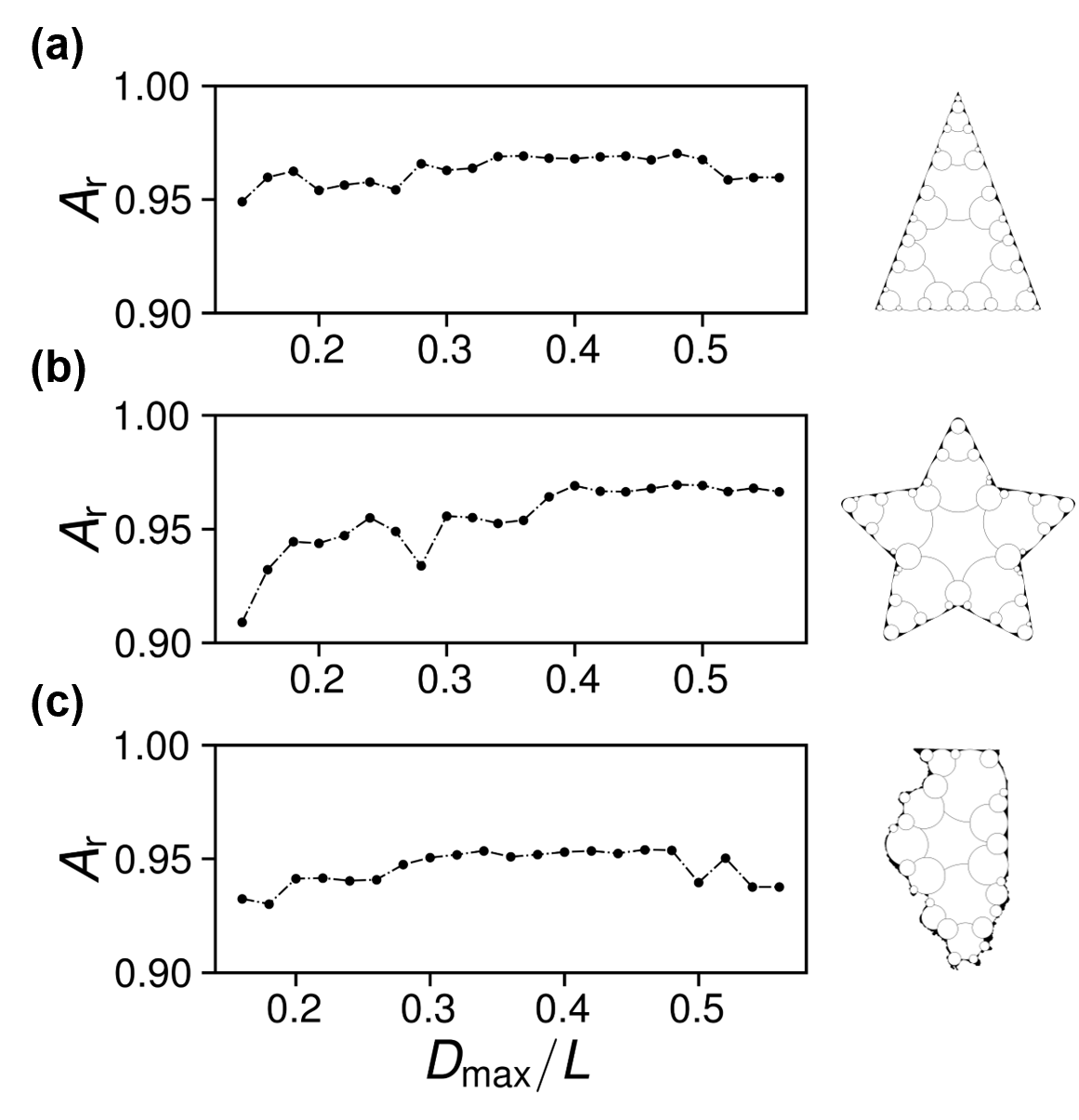}
\caption{(a,b,c) Area coverage ratio as a function of $D_{\rm max}/L$ for different target geometries, where $L$ is the larger of the horizontal and vertical dimensions of the target. A representative disk packing is shown to the right of each corresponding plot. Although the target shapes impose distinct geometric constraints through their boundary curvature, corners, and narrow regions, high coverage is achieved in all cases, demonstrating the flexibility of the algorithm across complex geometries. The fixed parameters are $v_{\rm f}=1.5~\mathrm{mm/s}$, $t_{\rm d}=0.1~\mathrm{s}$, and $N_{\rm i}=15$, matching those used in the experiments. The representative packings correspond to (a) $D_{\rm max}/L=0.48$ and $A_{\rm r}=0.97$,(b) $D_{\rm max}/L=0.40$ and $A_{\rm r}=0.96$, and (c) $D_{\rm max}/L=0.34$ and $A_{\rm r}=0.96$.}
\label{fig:MyFig7_Flexibility}
\end{figure}

\textit{Flexibility}. We finally show that the LGA can be used to generate a variety of target shapes. Fig.~\ref{fig:MyFig7_Flexibility} shows disk packings for several target geometries, along with their corresponding area coverage ratio as a function of $D_{\rm max}/L$, using $L=100~\mathrm{mm}$, $v_{\rm f}=1.5~\mathrm{mm/s}$, $t_{\rm d}=0.1~\mathrm{s}$, and $N_{\rm i}=15$. 
These process parameters are chosen to match the experimental setup used for the physical realization of the shapes described below. Although the kinetic constraints are fixed, the coverage curves differ across target shapes because each geometry imposes distinct packing constraints. Shapes with broad, smooth regions are more easily filled by large circular fronts, whereas sharp corners, narrow features, and highly curved boundaries require smaller disks and are therefore more sensitive to the discrete set of accessible radii. Nevertheless, area coverage ratios up to $A_{\rm r} = 0.95$ are achievable in all cases. Along with the experimental realizations presented in the next section, these results demonstrate that the algorithm provides an effective and reliable framework for generating complex shapes using isotropically expanding fronts.

\begin{figure*}[t]
\centering
\includegraphics[width=1\linewidth]{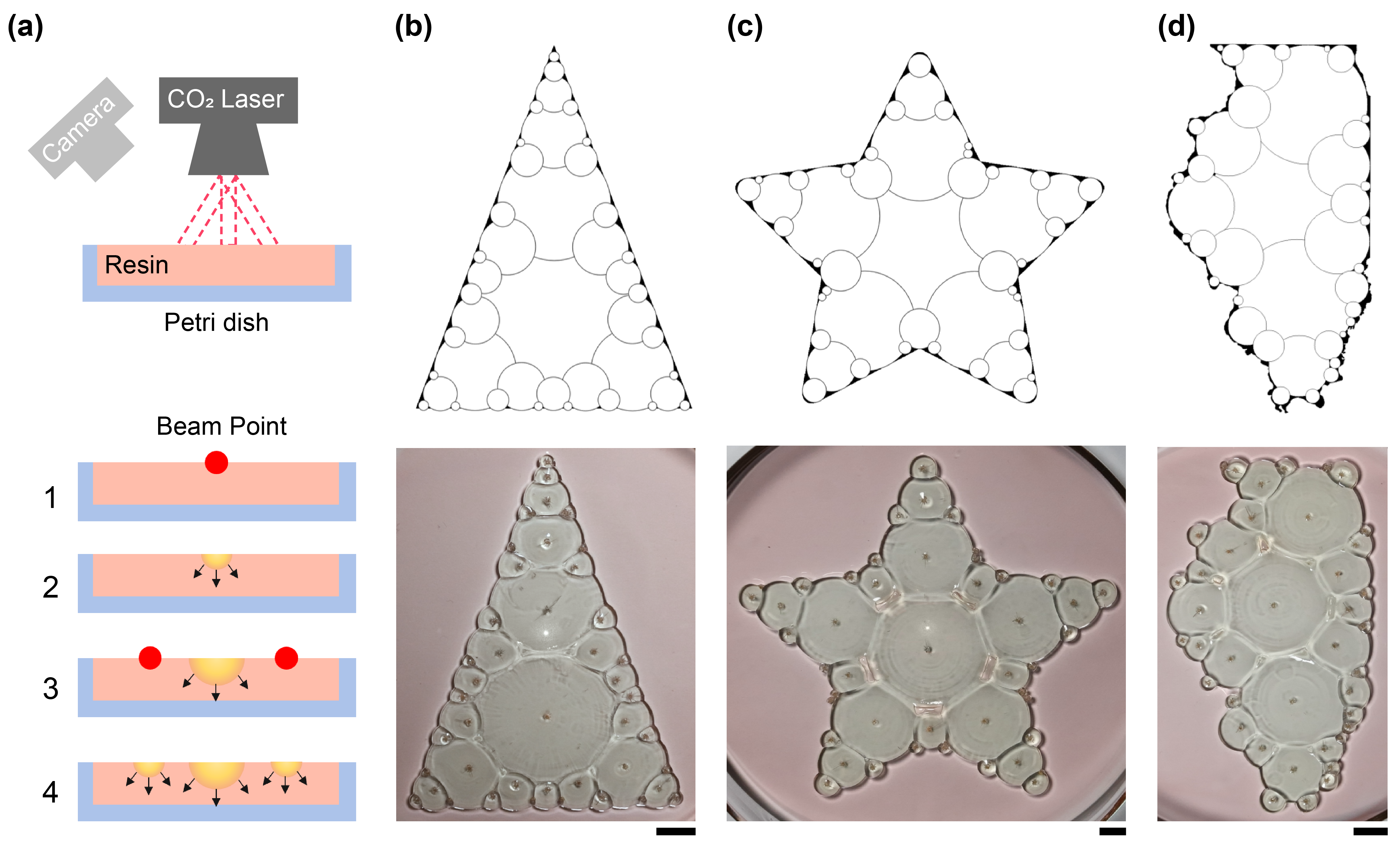}
\caption{(a) Schematic of the CO$_2$ laser setup (top) and dicyclopentadiene (DCPD) curing sequence (bottom):  1- Laser beam is irradiated to the surface of the resin, 2- Polymerization reaction starts from a point, 3- While initiated polymer curing continues to propagate, following CO$_2$ beams are irradiated to the surface starting new nucleation events, 4- The fronts continue to propagate until all resin in Petri dish is consumed. (b, c, d) Disk packings generated via LGA as shown in Fig.~\ref{fig:MyFig7_Flexibility} and their corresponding experimental realizations for (b) isosceles triangle, (c) star shape, and (d) map of Illinois with total growth times of $18~\mathrm{s}$, $16~\mathrm{s}$, and $15~\mathrm{s}$ respectively. All scale bars are $10~\mathrm{mm}$.}
\label{fig:MyFig8_Realization}
\end{figure*}

\section*{Experimental Realization}
We perform a proof-of-concept experimental realization of the LGA using frontal polymerization of dicyclopentadiene (DCPD). A viscosified DCPD resin (DCPD:catalyst:inhibitor ratio of $10{,}000:1:1$) is prepared as described by Lee et al.~\cite{lee2025multimaterial} and placed in a $140~\mathrm{mm}$ diameter Petri dish with a thickness of approximately $3~\mathrm{mm}$. The viscosity of the DCPD resin is tuned to about $5~\mathrm{Pa}\cdot\mathrm{s}$ to suppress excessive Marangoni flow at the surface, thereby facilitating stable initiation and front propagation.

The experimental setup consists of a commercial CO$_2$ laser system (Cloudray\texttrademark{} EC Series, $30~\mathrm{W}$) in which galvanometer-based motion provides scanning speeds up to $10{,}000~\mathrm{mm/s}$ in the $2$D plane. This rapid scanning allows a single laser beam to mimic the action of multiple initiators. Because of the high laser power and the thermal initiation requirements of frontal polymerization in DCPD, the beam can be redirected among different locations over a short time interval of approximately $10~\mathrm{ms}$. It can therefore cycle among as many as $15$ nucleation sites and repeatedly deposit heat at each site before significant thermal diffusion occurs. The sites are thus heated nearly simultaneously, allowing a single laser beam to effectively approximate the action of $15$ parallel initiators. 

A typical initiation condition uses $30~\mathrm{W}$ laser power with single-spot irradiation for $100$ loops at $57~\mathrm{kHz}$. Under these conditions, frontal polymerization is triggered at a spot in approximately $50~\mathrm{ms}$. For path planning, we use $t_{\rm d}=0.1~\mathrm{s}$ to reduce the sensitivity of the growth process to fluctuations in the reaction initiation time. Previous laser-triggered frontal polymerization experiments required the resin to be doped with carbon black because the near-infrared laser light ($800~\mathrm{nm}$) was not strongly absorbed by the neat resin, and initiation typically occurred on the order of $1~\mathrm{s}$~\cite{cook2024polymer}. By using a $10.6~\mathrm{\mu m}$ CO$_2$ laser, we directly heat the neat resin and initiate the front in less than $0.1~\mathrm{s}$, reducing the initiation time by an order of magnitude.

The output of the LGA, consisting of the disk-center positions and radii, is converted into a laser path plan by assuming a constant front velocity of $v_{\rm f}=1.5~\mathrm{mm/s}$. The CAD representation of the disk packing, along with the prescribed time intervals between consecutive initiation events, is then supplied to the CO$_2$ laser control software. The results for a few target shapes are shown in Fig.~\ref{fig:MyFig8_Realization}.

\section*{Conclusion}

In this work, we have shown that the growth of complex shapes can be modeled as the spatiotemporal packing of isotropically growing domains, which become disks in $2$D and spheres in $3$D. We introduced the Largest Gap Algorithm (LGA), which fills a target shape by assigning large disks to larger gaps and progressively smaller disks to finer features, while respecting the constraints that the growth kinetics impose on the accessible disk sizes. The resulting packing is not merely a geometric approximation, but a set of instructions that specifies where each front should initiate, when it should initiate, and how long each disk must grow.

A central result of this study is that these computational instructions can be realized physically. Using multi-point frontal polymerization of dicyclopentadiene (DCPD), we demonstrate that the computed initiation sequences can be translated into laser path plans and used to grow prescribed shapes experimentally. The final shape is formed not by tracing its entire boundary, but by coordinating a small number of local initiation events whose fronts expand, interact, and merge to produce the desired geometry.

Several open directions follow naturally from this work. The first is the extension from two dimensions to three. In principle, the same idea can be carried from disks to spheres, and from area coverage to volume filling. In practice, however, path planning becomes substantially more constrained in $3$D. In $2$D, a new front can be initiated anywhere at an exposed point in the plane. In $3$D, initiation sites may become buried by earlier growth events. Unless the remaining gaps remain accessible from the outer surface, they cannot be initiated later by an external heat source. Thus, $3$D growth is not simply a higher dimensional packing problem. It is also an accessibility problem in which the order of growth must preserve physical routes to the sites that remain to be activated.

A second direction is to broaden the class of elementary growth events. Here, we focused on point initiations, for which isotropic fronts grow into disks in $2$D and spheres in $3$D. If, however, the initial nucleation region is a line, a curve, or an extended seed of arbitrary shape, the resulting front need not be circular or spherical. A line initiation may grow into an oval or ovoid, while more general curves may generate asymmetric fronts. This raises a deeper question: what families of geometries can be grown when the building blocks are not only disks and spheres, but growing domains of varying shapes? Such generalized initiations may greatly expand the design space while preserving the speed and autonomy of frontal growth.

The broader promise of this approach is that it replaces point-by-point fabrication with an event-by-event autonomous growth process. Rather than drawing a shape along its perimeter or filling it line by line, one can initiate a distributed set of nucleation events and allow the material to complete the manufacturing by itself. Geometry is encoded in the timing and placement of nucleation points, while fabrication occurs through the natural propagation of the fronts. When the nucleation sites and initiation times are chosen intentionally, a complex shape can emerge in seconds. This possibility is striking because it describes a manufacturing process in which the fastest step is not the motion of a tool, but the release of stored chemical energy. The object is not assembled piece by piece, but is grown into its prescribed form using ideas from the packing of simple shapes.

\section{Acknowledgments}
We thank Philippe Geubelle for stimulating discussions on the topic throughout the project and for giving us early feedback on the manuscript. This work was supported by the U.S. National Science Foundation via grant AM350/NSF CMMI 24-52286. Computational resources were provided by NCSA Delta at UIUC through allocations MCH250097 and PHY240083 from the Advanced Cyberinfrastructure Coordination Ecosystem: Services \& Support (ACCESS) program, which is supported by U.S. National Science Foundation grants \#2138259, \#2138286, \#2138307, \#2137603, and \#2138296.

\bibliography{Refs}

\end{document}